\documentclass[aps,prl,preprint,superscriptaddress]{revtex4}
\usepackage{graphicx}
\usepackage{dcolumn}
\usepackage{bm}
\usepackage{amssymb}

\begin{document}

\title{Spin-state transition in LaCoO$_{3}$: direct neutron spectroscopic
evidence of excited magnetic states}

\author{A.~Podlesnyak}
\author{S.~Streule}
\author{J.~Mesot}
\affiliation{Laboratory for Neutron Scattering, ETH Z\"urich $\&$
Paul Scherrer Institut, CH-5232 Villigen PSI, Switzerland}
\author{M.~Medarde}
\author{E.~Pomjakushina}
\author{K.~Conder}
\affiliation{Laboratory for Developments and Methods, Paul
Scherrer Institut, CH-5232 Villigen PSI, Switzerland}
\affiliation{Laboratory for Neutron Scattering, ETH Z\"urich $\&$
Paul Scherrer Institut, CH-5232 Villigen PSI, Switzerland}
\author{M.~W.~Haverkort}
\author{D.~I.~Khomskii}
\affiliation{II~Physikalisches Institut, Universit{\"a}t zu K{\"o}ln,
Z{\"u}lpicher Strasse 77, 50937 K{\"o}ln, Germany}

\date{\today}

\begin{abstract}
A gradual spin-state transition occurs in LaCoO$_{3}$ around
$T\sim80-120$~K, whose detailed nature remains controversial. We
studied this transition by means of inelastic neutron scattering
(INS), and found that with increasing temperature an excitation at
$\sim0.6$~meV appears, whose intensity increases with temperature,
following the bulk magnetization. Within a model including crystal
field interaction and spin-orbit coupling we interpret this
excitation as originating from a transition between thermally
excited states located about 120~K above the ground state. We
further discuss the nature of the magnetic excited state in terms
of intermediate-spin (IS, t$_{2g}^{5}$e$_{g}^{1}$, $S=1$)
\textit{vs.} high-spin (HS, t$_{2g}^{4}$e$_{g}^{2}$, $S=2$)
states. Since the $g$-factor obtained from the field dependence of
the INS is $g\sim3$, the second interpretation looks more
plausible.
\end{abstract}


\maketitle

Due to its rich and in many respects puzzling properties,
LaCoO$_{3}$ keeps attracting attention and remains a controversial
topic. It is known that the ground state is nonmagnetic,
corresponding to a low-spin (LS) state of Co$^{3+}$ ions
(t$_{2g}^{6}$, $S=0$). However, with increasing temperature (as
well as with La$\rightarrow$Sr substitution) first a crossover
into a magnetic, but still insulating state appears at about
80-120~K, followed by another crossover into a "bad metallic",
magnetic state at $T\sim400-600$~K. The original interpretation of
the low-temperature crossover was done in terms of
thermally-induced population of the low-lying high-spin (HS) state
\cite{G1}; this process is furthermore favorized by thermal
expansion, since the HS Co$^{3+}$ has much larger radius
($\sim$0.75~\AA) than the LS state ($\sim$0.685\AA). Later,
especially after LDA+U band structure calculation have become
available \cite{Korotin}, another interpretation was put forward:
within this scenario, the first crossover at $\sim100$~K would be
due to a transition into an intermediate-spin (IS) state. This
interpretation was supported by a number of experimental evidences
\cite{Masuda,Saitoh,Abbate,Yam2,Ishi,Louca1,Louca2, Zobel, Maris}.
None of these arguments however gave a definite proof that the
first thermally-excited state is indeed the IS one. Very recent
measurements indicate that the first excited state could still be
the HS state \cite{Noguchi, Ropka, Kobayashi, Kyomen1, Kyomen2}.

Also theoretically the situation is not clear: Hartree-Fock calculations showed
that the HS state, or the HS-LS ordered state is more stable than the IS state
\cite{Zhuang}, in contrast to LDA+U calculations \cite{Korotin}. Thereby, model
calculations on a CoO$_6$ cluster explicitly including the Co-O hybridization
can not reproduce an IS ground-state \cite{Haverkort}, indicating that the
proposed mechanism why LDA+U finds an IS as first excited state, namely large
covalency, is rather questionable.

With this controversy in mind, we undertake a neutron scattering
study of LaCoO$_{3}$ at different temperatures with the goal of
identifying the energy level of the thermally excited state of
Co$^{3+}$. Indeed, we discovered that a rather unusual feature in
the spectrum appears with increasing temperature in forms of
thermally-induced relatively sharp inelastic peak at an
energy-transfer of $\sim0.6$~meV \cite{Louca3}. The intensity of
this peak strongly increases with $T$ following the behavior of
the magnetic susceptibility $\chi(T)$, suggesting that the
inelastic scattering occurs between thermally populated magnetic
states of LaCoO$_{3}$. The position and the temperature dependence
of the intensity of this peak also coincide with the excitations
observed in LaCoO$_{3}$ by ESR \cite{Noguchi, Kataev}. By
analyzing the features of this novel excitation, and combining it
with model calculations, we discuss the two possible scenarios
mentioned above.

Polycrystalline LaCoO$_{3}$ was prepared by standard sintered
techniques using La$_2$O$_3$ and Co$_3$O$_4$ of a minimum purity
of 99.99\%. The respective amounts of starting reagents were mixed
and calcinated at temperatures 1000-1200$^{\circ}$~C during at
least 100~h in air, with several intermediate grindings. The
sample was checked by x-ray diffraction and found to be single
phase within experimental accuracy. The space group $R\bar{3}c$
and lattice parameters of $a=5.4433(1)$~\AA, $c=13.0932(4)$~\AA
~are in agreement with previously published data
\cite{G1,Asai1,Thornton}. The inelastic neutron scattering (INS)
measurements were performed on the time-of-flight spectrometer
FOCUS \cite{Mesot} installed at the spallation neutron source SINQ
at Paul Scherrer Institut, Villigen, Switzerland. Zero field
experiments were carried out in the temperature interval of
$1.5-100$~K using a conventional helium cryostat. The data were
collected using incoming neutron energies of 3.5 and 20~meV,
giving an energy resolution at the elastic position [full width at
half maximum (FWHM)] of 0.1 and 1.6~meV, respectively. The
triple-axis spectrometer TASP with final neutron energy 4.7~meV
was used for the measurements in external magnetic field up to
$H=6$~T.

\begin{figure}[tb!]
%\begin{center}
\includegraphics[width=0.65\textwidth]{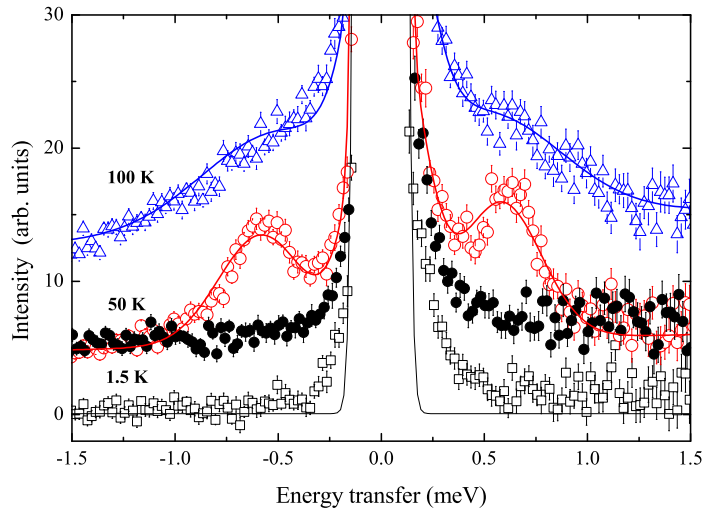}
\caption {(Color online) Temperature evolution of the INS profiles
measured in LaCoO$_{3}$. The filled circles correspond to the LaAlO$_{3}$
nonmagnetic reference compound at $T=50$~K.
The lines are the result of least-squares fits using Gaussian
functions to describe the lineshape of the transition. For clarity, an offset
has been added to the various curves.
\label{INS}}
%\end{center}
\end{figure}

The high-resolution low-energy transfer inelastic spectra for a few selected
temperatures are shown in Fig.~\ref{INS}. There are no excitations in the energy
window $E<1.5$~meV for temperatures $T<30$~K. A single inelastic peak at
an energy transfer $ \delta E=0.61 \pm 0.05$~meV was found at intermediate
temperatures starting from $T\sim30$~K. A strong broadening of the transition was
observed with increasing temperature. Note that the spectra obtained from a
non-magnetic reference compound, LaAlO$_{3}$, remain structureless at all temperatures.
The high-energy transfer spectra observed for LaCoO$_{3}$ exhibit several broad
inelastic peaks at about 10, 14 and 22~mev (not shown). However, all these peaks
exhibit clear increase of their intensity with scattering vector and
temperature. Therefore we conclude that they are due to phonon scattering, in
agreement with previously published data \cite{Louca2}. No evidence for other
magnetic excitations was observed in the LaCoO$_{3}$ spectra within the available
energy window.

\begin{figure}[tb!]
%\begin{center}
\includegraphics[width=0.65\textwidth]{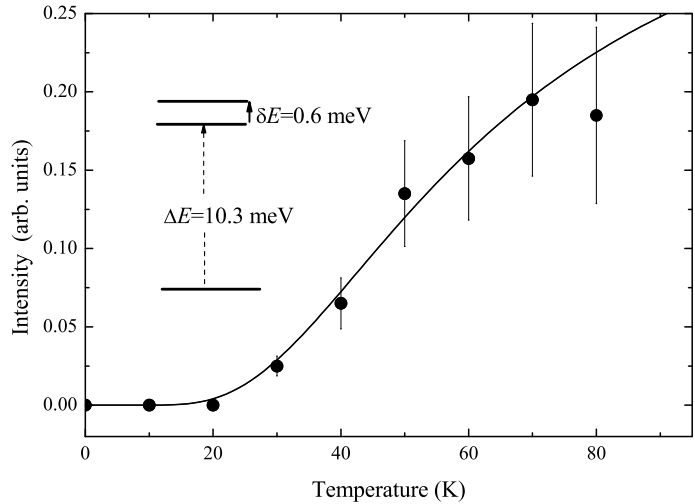}
\caption {Observed (circles) and fitted (solid line) temperature
dependence of the integral intensity of the INS peak at 0.6~meV.
The resulting level scheme is shown in the inset.
\label{Temp_dep}}
%\end{center}
\end{figure}

For  noninteracting ions the thermal neutron cross section for the
transition $|\Gamma_i\rangle \rightarrow |\Gamma_j\rangle$ is
given in the dipole approximation by \cite{Marshall}
\begin{widetext}
\begin{eqnarray}
\frac{d^2\sigma}{d\Omega d\omega} \sim \frac{1}{Z} \exp \Big(
-\frac{E_i}{k_\textsc{b}T} \Big) F^2(\mathbf{Q})
|\langle\Gamma_j|J_\perp|\Gamma_i\rangle|^2 \delta(E_i - E_j \pm \hbar
\omega)~. \label{eqcros}
\end{eqnarray}
\end{widetext}
Here $F^2(\mathbf{Q})$ is the magnetic form factor, $J_\perp$ is
the component of the total angular momentum operator perpendicular
to the scattering vector $\mathbf{Q}$, and $Z$ is the partition
function. It follows from Eq.~(\ref{eqcros}) that the energy gap
$\Delta E=E_{i}-E_{0}$ between the ground-state and the
excited-state can be deduced either directly from the position of
the corresponding inelastic peak (in case of nonzero matrix
element $|\langle\Gamma_0|J_\perp|\Gamma_i\rangle|$), or from the
temperature dependence of the transition between two excited
levels $\delta E=E_{j}-E_{i}$ which is governed by Boltzman
statistics. Note that the direct transition $\Delta E$ out of the
ground-state was observed neither in the previous INS experiments
\cite{Asai2,Louca2}, nor in our current measurements, most likely
due to selection rules. Therefore, in order to determine the
energy of the excited state we apply the least-squares fitting
procedure to the temperature dependence of the integrated
intensity $I$ of the INS signal as shown in Fig.~\ref{Temp_dep}.
The position of the excited states turns out to be $10.3 \pm
1$~meV, which coincides well with the results obtained from ESR
(12~meV, ref.~\onlinecite{Noguchi}). Our estimation is based on a
temperature independent level splitting scheme. Although we can
not exclude a slight variation of the position of the excited
states due to thermal expansion of the unit cell, a level crossing
of the ground- and excited states as suggested from the LDA+U
calculations \cite{Korotin} can be excluded, since this would
result in a non-monotonic temperature behavior of $I$ around
$T<80$~K, which was not observed in our experiment. Furthermore,
the position of the peak at $\delta E=0.6$~meV is unaltered,
suggesting that the trigonal CF remains nearly constant in this
temperature range. Thus, our results imply that the first broad
peak in magnetic susceptibility at $\sim100$~K is due to a gradual
thermal population of the excited levels rather than a
modification of the level scheme due to a phase transition.

\begin{figure}[tb!]
%\begin{center}
\includegraphics[width=0.65\textwidth]{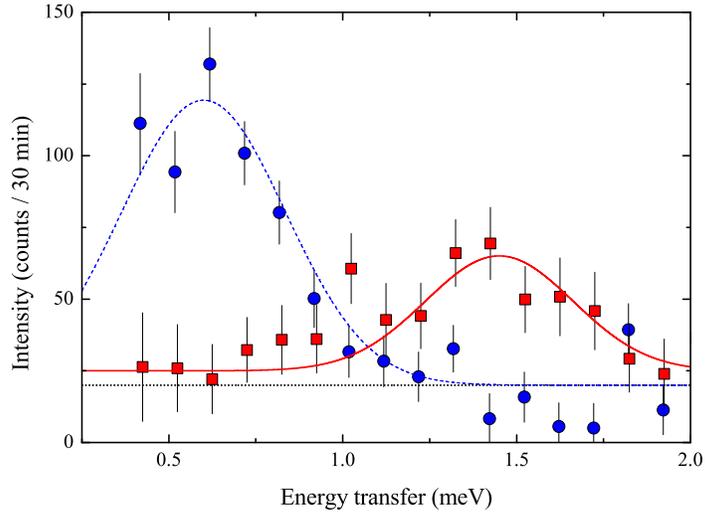}
\caption {(Color online) The magnetic inelastic scattering at
$T=50$~K in 0~T (circles) and 6~T (boxes) applied field.
\label{H_dep}}
%\end{center}
\end{figure}

The observed magnetic INS, which was obtained as the difference of
the intensities at 50 and 5~K, is shown in Fig.~\ref{H_dep}. A
clear shift of the transition to the higher energy $\sim 1.5$~meV
was observed in magnetic field $H=6$~T compared to the zero-field
spectrum, thus firmly establishing its magnetic origin. Due to
weakness of the signal we can not conclude whether the peak is
split in the external magnetic field. The change in energy of this
peak from $0.6$ meV to about 1.5~meV in magnetic field of 6~T is
in good agrement with the $g$-factor measured from ESR experiments
\cite{Noguchi}.

\begin{figure}[tb!]
%\begin{center}
\includegraphics[width=0.65\textwidth]{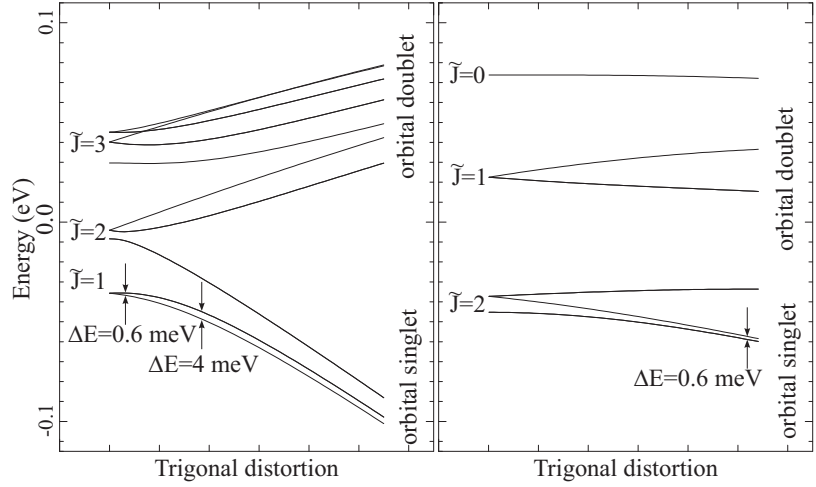}
\caption {Schematic diagram of the excitation spectrum of
high-spin (left) and intermediate-spin (right) Co$^{3+}$ as a
function of trigonal distortion (see text). \label{enlev}}
%\end{center}
\end{figure}

Let us discuss the possible origin of this excited state. There
are two possibilities: either high-spin or a intermediate-spin
states of Co$^{3+}$. The HS state with $S=2$ has, in a cubic CF,
the occupation t$_{2g}^{4}$e$_{g}^{2}$, i.e., it has half-filled
shell of t$_{2g}^{3}$e$_{g}^{2}$, say with spins up, and one extra
spin-down electron on a triple-degenerate $t_{2g}$-level, which
can be described by an effective orbital moment $\widetilde{L}=1$
\cite{Abragham}. Total multiplicity of this state is
$(2S+1)(2\widetilde{L}+1)=15$. Spin-orbit coupling splits this
state into the lowest-lying triplet $\widetilde{J} = S -
\widetilde{L} = 1$, next is a quintet $\widetilde{J} = 2$, and the
highest-lying is state has $\widetilde{J} = 3$. If the system is
strongly distorted there will be a ground-state orbital singlet
and a higher excited orbital doublet. In the left panel of Fig.~
\ref{enlev} we show the energy level diagram for the high-spin
state as a function of trigonal distortion. This energy level
diagram has been calculated for a CoO$_6^{9-}$ cluster, including
full multiplet theory, spin-orbit coupling and Co-O hybridization.
For the Slater integrals and the spin orbit coupling atomic
Hartree-Fock values are used, the hopping parameters are according
to Harrisons's rules \cite{Harrison}. The calculations have been
done with the use of the program XTLS8 \cite{Tanaka}. For the HS
there are two places in the energy-level diagram, where an
excitation of 0.6 meV can happen. If the trigonal crystal field is
relatively small, the $\widetilde{J} = 1$ triplet will be split by
this crystal field. On the other-hand, if it is rather large, the
orbital singlet with $S=2$ will be split due to spin-orbit
coupling. In both cases the splitting is governed by second order
effects and the resulting splitting is much smaller than the
perturbing interaction. The scenario of a small crystal-field with
respect to the spin-orbit coupling has been discussed in quite
details recently \cite{Ropka}. The scenario of a large
crystal-field with respect to the spin-orbit coupling is
equivalent to a spin only scenario.

One should also consider what would be the situation if the first
excited state is an IS Co$^{3+}$, which follows from LDA+U
theoretical calculations \cite{Korotin} and which was used to
interpret a number of experimental data
\cite{Masuda,Saitoh,Abbate,Yam2,Ishi,Louca1,Louca2, Zobel, Maris}.
This case is actually much more interesting and more difficult to
treat theoretically. First thing to note is that in this case we
have one electron in the $e_{g}$-shell and one hole ($t_{2g}^{5}$
occupation) in the $t_{2g}$-shell. Their Coulomb interaction
strongly depends on the particular orbital occupation. The
$(x^2-y^2)$-electron has strong attraction to the $(xy)$-hole, so
that $(x^2-y^2)^1(\underline{xy})^1$ state has much lower energy
than e.g. $(z^2)^1(\underline{xy})^1$. Thus we can consider the
lowest states on a basis of $ (x^2-y^2)^1(\underline{xy})^1,
~(x^2-z^2)^1(\underline{xz})^1 ~\textrm{and}~
(xy^2-z^2)^1(\underline{yz})^1 $ \cite{Haverkort}. Therefore the
total orbital degeneracy of the IS state in a cubic CF is 3 and
not 6 $(3t_{2g} \times 2e_{g})$ as one could have expected. We can
thus again describe these states by the effective orbital triplet
$\widetilde{L}=1$; but because of the more complicated type of the
basis states, the maximum magnitude of the magnetic moment is not
1 but $\frac{1}{2}$ (i.e. $L_z=\{ \frac{1}{2}, 0, -\frac{1}{2} \}$
\footnote{One can see this by expressing nonorthogonal
$(x^2-y^2)$, $(x^2-z^2)$ and $(y^2-z^2)$ through an orthogonal
basis $(x^2-y^2)$, $z^2$, so that
$(x^2-z^2)/(y^2-z^2)=\pm\sqrt{\frac{1}{4}}
(x^2-y^2)-\sqrt{\frac{3}{4}}z^2$}.). In other words, one has an
effective orbital $g$-factor of $\frac{1}{2}$. As a result, we are
dealing with 9 states $(2\widetilde{L}+1) (2S+1)$ (with $S=1$ for
IS state), which are split by the spin-orbit coupling into
multiplets with $\widetilde{J}=2,1,0$. However, in this case the
quintet $\widetilde{J}=2$ is the lowest state. Thus, the
multiplicity of the IS state in cubic CF is 5. Strong enough
distortions, or orbital ordering, modify the energy-level scheme
as shown in the right panel of Fig. \ref{enlev}. If the
distortions are larger than the spin-orbit coupling constant, the
ground-state becomes an orbital singlet. This orbital singlet is
split due to second-order spin-orbit interactions into two levels
that could very well be 0.6 meV apart from each other. In cubic
symmetry, the $\widetilde{J}=2$ quintet originating from the IS
state is however also split due to second order interactions. This
splitting seems to be somewhat larger than the measured value of
0.6 meV.

If we now compare our results with other measurements, we notice
that from a comparison of the magnetic susceptibility with the
anomalous expansion coefficient of LaCoO$_{3}$, Zobel \textit{et
al.} \cite{Zobel} concluded that the degeneracy of the first
magnetic excited state is 3. This leaves only two possibilities
open. The first magnetic excited state in LaCoO$_{3}$ can be a HS
state with a small non-cubic crystal field, or it can be an IS
state with a large non-cubic crystal field or orbital ordering.
There is one striking difference between these scenarios: this is
the predicted $g$-factor. The HS state with a small distortion is
a triplet with a $g$-factor of about 3.5 \cite{Abragham, Ropka},
whereas the IS with strong distortion is a triplet with a
$g$-factor of about 2.0. ESR measurements found a $g$-factor of
3.35 - 3.55 \cite{Noguchi}, supporting the HS state; this also
agrees with our results since we obtained $g\sim3$ (see
Fig.~\ref{H_dep}). On the other hand a fit to the magnetic
susceptibility yields a $g$-factor of about 2.28 \cite{Zobel},
supporting the IS scenario.

To summarize, we observe a novel inelastic excitation in
LaCoO$_{3}$ which is due a thermally excited magnetic state of
Co$^{3+}$ ions. This confirms the presence of  thermally induced
spin-state transition (or rather crossover) at $T\sim100$~K from
the LS Co$^{3+}$ to a magnetic HS or IS state. We discuss both
possibilities theoretically and show that one can explain both our
and other results (thermodynamic, ESR) in the framework of a
HS-triplet as first excited state, with the $g$-factor $\sim3.5$,
weakly split by small distortions from the cubic symmetry. Another
possibility would be the IS state with orbital ordering or strong
non-cubic crystal-fields, which however would result in a
spin-only system with a $g$-factor $\sim2.0$, a value difficult to
reconcile with our experimental data. Thus the first
interpretation (HS excited state) seems to us more plausible.

\begin{acknowledgments}
The experiments have been performed at the Swiss Spallation Neutron Source,
Paul Scherrer Institute, Villigen, Switzerland. We are indebted to the Swiss
National Science Foundation for financial support through grant 200021-100194
and NCCR MaNEP project. Financial support by the Deutsche Forschungsgemeinschaft via SFB 608
and by the  Marie Heim-V{\"o}gtlin program
(grant No. PMPD2--102504) is also gratefully acknowledged. We would like to
thank Arata Tanaka for the kind use of his program XTLS8.
\end{acknowledgments}

%\bibliographystyle{apsrev}
%\bibliography{LaCoO3new_04}

\end{document}